\documentclass[twocolumn,twoside]{article}

\usepackage{a4}
\usepackage{amssymb}
\usepackage{amsmath}
\usepackage[numbers,sort&compress]{natbib}
\usepackage{graphicx}
\usepackage{blindtext}

\begin{document}

\title{\bf {\Large Synchrotron radiation in nonuniform magnetic field}} 

\date{{\normalsize\textit{$^{1}$Lebedev Physical Institute, Russian Academy of Sciences,
Leninskii prosp. 53, 119991, Moscow, Russian Federation,\\
$^{2}$Moscow Institute of Physics and Technology
(National Research University), Institutskii per. 9,
141701 Dolgoprudny, Moscow region, Russian Federation}} \\[1ex]
{\small \textit{Usp.\ Fiz.\ Nauk} \textbf{196}(4), 
 436--445 (2026) 
[in Russian]\\
English translation: \textit{Physics -- Uspekhi}, \textbf{69} (4) 406--414 (2026)}
\\{\small Translated by the authors}
}

\author{V.S.~Beskin$^{(1,2)}$, A.Yu. ~Istomin$^{(2,1)}$, F.A.~Kniazev$^{(2,1)}$, T.I.~Khalilov$^{(2,1)}$}

\maketitle

{\bf {\underline {Abstract.}} The distortion of the spectrum of synchrotron radiation received by a distant observer due to magnetic field inhomogeneity is discussed.
It is shown that for single-particle radiation, a noticeable difference in the integrated spectrum from the case of a uniform magnetic field
occurs only for observation times significantly exceeding the observer's departure from the beam pattern, and only for subrelativistic
particles, when the radiation arrives from regions with significantly different magnetic fields during the observation time. For shorter
observation times, field inhomogeneity leads only to small-scale distortions (smaller or comparable in frequency to the gyrofrequency
$\omega_{\rm B}$) and does not manifest itself in the averaged spectrum. For the case of a particle ensemble, which is of interest for astrophysical
applications, the difference in the integrated spectrum becomes negligible.}


\setcounter{secnumdepth}{3}
\setcounter{tocdepth}{2}

\tableofcontents

\section{Preface}
\noindent

Remembering Sergei Ivanovich Syrovatskii in the year of his centenary (one of the authors of this note was his last student at the Department of Problems of Physics and Astrophysics at Moscow Institute of Physics and Technology in the late 1970s), one cannot help but be amazed at the lasting memory he left behind. Sergei Ivanovich passed away almost fifty years ago, but throughout these years, seminars, conferences~\cite{mem1}, and sessions of the Physical Sciences Division of the Russian Academy of Sciences~\cite{mem2, mem3, mem4, mem5, mem6, mem7, mem8, mem9, mem10, mem11, mem12, mem13} dedicated to his memory have been held continuously.

And here, of course, we are not just talking about his scientific achievements. Certainly, his personal qualities, such as integrity and benevolence, played a significant role, greatly influencing the development of an atmosphere of high scientific and moral standards, without which the golden age of Soviet astrophysics would not have been possible. Sergei Ivanovich always championed the principles of scientific integrity and respect for the work of others --- principles he always followed unwaveringly.

Another of Sergei Ivanovich's most important qualities was his profound understanding of the very essence of physical research. When discussing a scientific article or student work at a seminar (he had his own weekly seminar on magnetohydrodynamics!), Sergei Ivanovich always emphasized the accuracy (or, conversely, the inaccuracy) of the formulation of the physical problem and the boundary conditions. S.I.~Syrovatskii believed that exact solutions, even of approximate equations, are extremely important for developing our intuition, which subsequently allows us to qualitatively understand the fundamental properties of ongoing processes. Here, in particular, his affiliation with the school of I.E.~Tamm was evident, arguing that any interpretation of observations must be based on fundamental physics. Incidentally, on this issue, he argued with Ya.B.~Zeldovich, who, on the contrary, believed that it was necessary to primarily focus on the analysis of approximate solutions of exact equations.

All this enabled Sergei Ivanovich to work at the very forefront of the then rapidly developing theoretical astrophysics. His most important results were in magneto-hydrodynamics (classification of discontinuities and shock waves, the question of their evolutionary nature, analysis of the stability of tangential discontinuities), radio astronomy (theory of cosmic synchrotron radiation taking into account the non-uniform distribution, diffusion, and energy loss of electrons), cosmic ray astrophysics (questions of exceptional nuclear acceleration and the universality of their spectra), and solar physics. The fact that this note was not conceived in connection with his anniversary once again demonstrates the ultimately fundamental role Sergei Ivanovich Syrovatskii's research played in the development of modern astrophysics.

\section{Introduction}
\noindent

As already noted, one of S.I.~Syrovatskii's areas of research was the theory of synchrotron radiation --- one of the most important foundations of modern physics. In particular, it is one of the key tools of astrophysics, allowing us to understand the processes occurring in cosmic plasma. Many discoveries can be cited here, such as, the explanation of the non-thermal (synchrotron) optical radiation of the Crab Nebula~\cite{G, S}. A detailed exposition of the theory of synchrotron radiation can be found in the articles by V.L.~Ginzburg and S.I.~Syrovatskii~\cite{Syrov}, V.L.~Ginzburg, V.N.~Sazonov, and S.I.~Syrovatskii~\cite{Sazonov}, published in Phys. Uspekhi in 1965--1968 (the results of which were later included in reviews and monographs~\cite{GinSyr, ARAA, Ginzburg}), as well as, for example, in A.G.~Pacholczyk's book ''Radio Astrophysics'' ~\cite{Pach} published in Russian in 1973.

As is well known, the theory of synchrotron radiation is based on the assumption of a uniform magnetic field, which, for a constant energy $m_{\rm e}c^2\gamma$ ($\gamma \gg 1$) of the emitting particle, leads to a periodic signal ${\bf E}({\bf r}, t)$~\cite{LL}
\begin{equation}
{\bf E}({\bf r},t) = \frac{e}{c^2 r}\frac{[{\bf n}\times [({\bf n} - {\bf v}/c)\times {\bf w}]]}{(1 - {\bf nv}/c)^3},
\label{E}
\end{equation}
received by the observer. Here ${\bf n}$ is the unit vector in the direction of radiation, while ${\bf v}$ and ${\bf w}$ are, respectively, the velocity and acceleration of the particle at the retarded moment of time $t^{\prime} = t - r^{\prime}\left(t^{\prime}\right)/c$, where $r^{\prime}$ is the distance from the charged particle to the observer. Here, the fundamental frequency $\omega_{0}$, the harmonics of which determine the spectrum of the observed radiation, has the form
\begin{equation}
\omega_{0} = \frac{\omega_{B}}{1 - \beta_{\parallel}\cos\Theta},
\label{omega}
\end{equation}
where, by definition,
\begin{equation}
\omega_{B} = \frac{eB}{\gamma m_{\rm e}c}
\label{omegaB*}
\end{equation}
is the relativistic gyrofrequency, $\Theta$ is the observation angle, and $\beta_{\parallel} = v_{\parallel}/c = \beta\cos\chi$, where $\chi$ is the pitch angle of the emitting particle.

As a result, due to the harmonic motion of the radiating particle along a spiral, it is possible to quite simply expand the electric field ${\bf E}(r, t)$ into a Fourier series~\cite{Sazonov}
\begin{eqnarray}
{\bf E}_{n}(\Theta, r) = \frac{2e}{cr}\,\frac{n\omega_{B}\beta\sin\chi}{(1 - \beta\cos\chi\cos\Theta)^2}\,
\nonumber \\
\times \left[{\bf l}_{1}J^{\prime}_{n}(z_{n}) - i{\bf l}_{2}\frac{\cos\Theta - \beta\cos\chi}{\beta\sin\chi\sin\Theta}J_{n}(z_{n})\right].
\label{eq:En}
\end{eqnarray}
Here
\begin{equation}
z_{n} = n\frac{\beta\sin\chi\sin\Theta}{1 - \beta\cos\chi\cos\Theta},
\label{zn}
\end{equation}
the unit vector ${\bf l}_{2}$ is directed along the magnetic field in the picture plane, and \mbox{${\bf l}_{1} = {\bf l}_{2}\times {\bf n}$.} Accordingly, one can also find the spectral radiation power at a given harmonic in the direction of the observer
\begin{equation}
{\tilde p}_{n}(\Theta, r) = \frac{c}{4 \pi} |{\bf E}_{n}(\Theta, r)|^2.
\label{pn}
\end{equation}
The main energy of the radiation will be contained in high harmonics $\omega_{n} = n\,\omega_{0}$ with $n \sim \gamma^{3}$, so that for the spectrum averaged over all angles $\Theta$, the maximum of the spectrum corresponds to the frequency {\footnote{For spectral density  ${\tilde p}_{n}(\Theta, r)$ (\ref{pn}) this is not the case, see the figure below.~\ref{fig2}.}}
\begin{equation}
\omega_{\rm max} =  0.44 \, \omega_{B}\gamma^3.
\label{omegamax}
\end{equation}
This frequency, as is known, is directly related to the characteristic duration time of the radiation burst $\tau$ registered by the observer ($\omega_{\rm max} \approx 2\pi/\tau$). As for the synchrotron radiation spectrum, for high harmonics one can switch to a continuous spectrum
\begin{equation}
{\tilde p}(\omega, \Theta) = {\tilde p}_{n}(\Theta)\frac{n}{\omega_{n}}.
\label{pom}
\end{equation}
The result is textbook relationships expressed in terms of the Macdonald functions $K(x)$. For example, for the total spectral power loss, i.e., for the quantity ${\cal J}(\omega) = 2\pi r^2\int{\tilde p}(\omega, \Theta)\sin\Theta {\rm d}\Theta$ (${\rm d}I = {\cal J}(\omega){\rm d}\omega$), we have~\cite{LL}
\begin{equation}
{\cal J}(\omega) = \frac{\sqrt{3}}{2\pi}\, \frac{e^2 \gamma \omega_{B}}{c} \xi \int_{\xi}^{\infty}K_{5/3}(\xi') {\rm d}\xi',
\label{eq:integrated_spectrum}
\end{equation}
where $\xi = \omega/\omega_{\rm c}$, and
\begin{equation}
\omega_{\rm c} = \frac{3}{2}\omega_{B}\gamma^3.
\label{omegac}
\end{equation}

Indeed, there have been attempts to generalize the corresponding expressions to the case of an inhomogeneous magnetic field~\cite{Thorne, Cavallo, Axford, Medvedev}. In particular, Medvedev~\cite{Medvedev} investigated the so-called jitter radiation emitted by the charged particles, moving in a highly inhomogeneous small-scale magnetic field. Much attention has also been paid to the so-called synchro-curvature radiation~\cite{ChZh, Vagano, Kelner}, when the particle motion is directed practically along the curved magnetic field, and the radiation itself is described by an intermediate regime, which, depending on the curvature radius of the field lines, the gyroradius (Larmor radius), and the pitch angle, transforms into the limiting cases of purely synchrotron or purely curvature radiation, respectively. In this case, the estimates usually did not go beyond the well-known synchrotron formulas, in which the parameters of the medium depended on the coordinates (see, for example,~\cite{YuAKovalev}). This approach, however, is not self-consistent.

Indeed, let us turn to the simplest geometry of a large-scale inhomogeneous (namely, monopole) magnetic field ${\bf B}({\bf r}) = B_{0}(R/r)^{2}\,({\bf r}/r)$ and consider a radiating particle whose guilding center moves along the $z$-axis. In turn, we place the observer at infinity at an angle $\Theta$ to this axis. In this case, for which the equations of motion can be easily integrated (see Appendix) and, therefore, the time dependence of the electric field ${\bf E}(t)$ (\ref{E}) is determined, the fundamental frequency (\ref{omega}), which determines the spectrum of the observed radiation, in the approximation $v = c$ takes the form
\begin{equation}
\omega_{0}^{*} = \frac{\omega_{B}R^2}{r^2(1 - \beta \cos\chi\cos\Theta)},
\label{omegar}
\end{equation}
where again $\chi$ is the pitch angle of the emitting particle, and $R$ is the distance at which the frequency $\omega_{B} = e B(R) / \gamma m_ec$ is determined (it is necessary that the gyroradius of the particle be much smaller than all the distances under consideration). A new crucial element is that, due to the conservation of the first adiabatic invariant $p_{\perp}^2/B$, the angle $\chi$ will change over time, while preserving the product
\begin{equation}
r\sin\chi = {\rm const}.
\label{rsinchi}
\end{equation}

\begin{figure}
\centering
\includegraphics[scale=0.4]{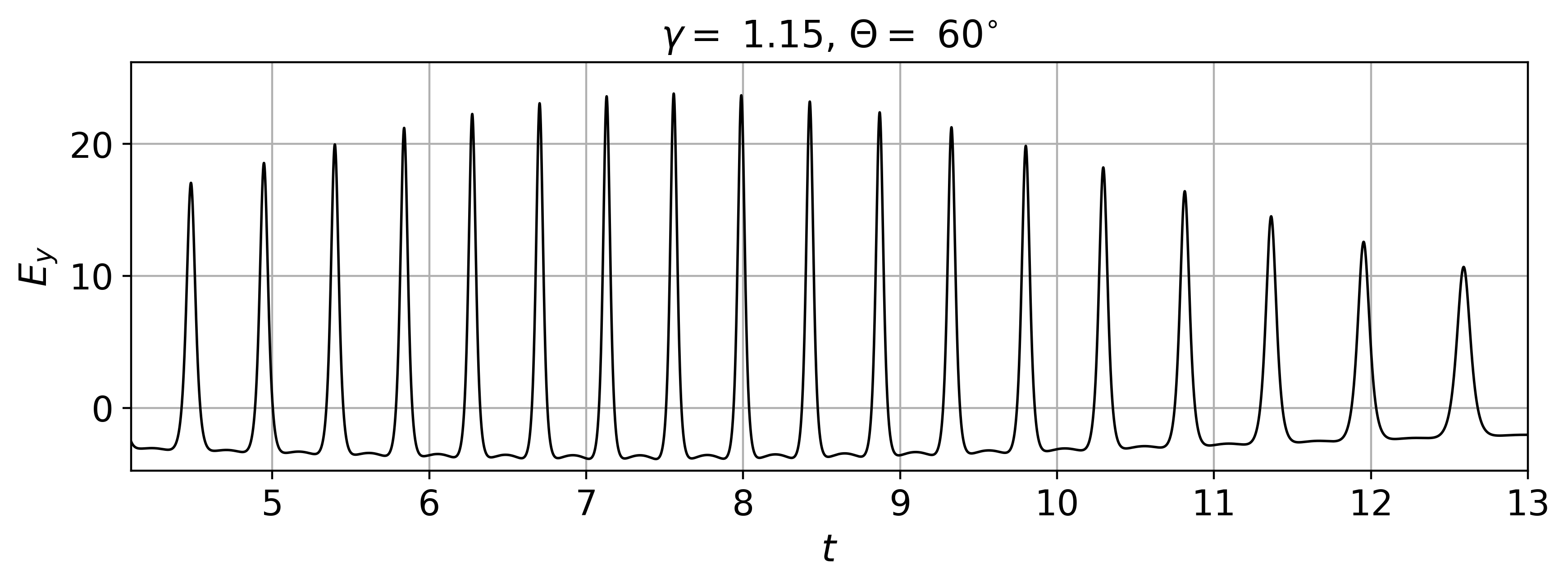}
\caption{The electric field $E_{y} = ({\bf l}_{1}\cdot{\bf E}(t))$ (\ref{E}), measured by an observer located at infinity at an angle $\Theta$ to the $z$-axis (in arbitrary units). The parameters are chosen such that changes in the pulse width and repetition rate are visually noticeable.}
\label{fig1}
\end{figure}

As a result, as can be seen from relation (\ref{omegar}), the received signal ceases to be periodic --- the distance between the radiation peaks, as shown in Fig.~\ref{fig1}, gradually increases. It should be emphasized here that we are talking specifically about a signal received by an observer located at an infinitely large distance at an angle $\Theta$ to the $z$ axis. If we consider the radiation frequency along the $\Theta = \chi$ direction, then this frequency, thanks to the relation (\ref{rsinchi}), would be constant in the limit $v \rightarrow c$. In fact, this very formulation was considered in the monograph~\cite{Ginzburg}, in which many formulas (including the expression for ${\bf E}_{n}$) are written out after the substitution $\Theta = \chi$ and $v = c$.

Furthermore, even ignoring the change in the fundamental frequency, we see that changing the pitch angle relative to the observation angle leads to two more effects: changes in the amplitude and duration of the bursts. Clearly, the maximum amplitude and minimum duration of the bursts, $\tau_{n}$, are reached precisely at the moment when $\chi = \Theta$. Furthermore, as the burst amplitude decreases, its duration also increases.

Finally, in contrast to the standard formulation of the problem, when (ignoring the change in the energy of the emitting particle) the radiation in a given direction exists infinitely long, in our problem, as is also clearly seen in Fig.~\ref{fig1}, the observer will enter the radiation pattern of the emitting particle only for a finite time $t_{\rm tot}$. It can be easily estimated, knowing that the radiation occurs at the angle $\vartheta \sim 1/\gamma$. Now using the relation $\vartheta \sim \delta r/r$, following from (\ref{rsinchi}) and the estimate $t_{\rm tot} \sim \delta r/c$, we finally obtain $t_{\rm tot} \sim r/(\gamma c)$. This circumstance should also be taken into account.

It would seem that all the factors listed above should have led to a significant change in the observed synchrotron radiation spectrum. However, as will be shown below, the field inhomogeneity leads only to small-scale distortions, with virtually no effect on the integrated spectrum. This note is devoted to a detailed explanation of this interesting property.

\section{Small-scale distortions}
\noindent

One of the applications that led to the consideration of the corrections to the synchrotron radiation associated with magnetic field inhomogeneity was the question of the radiation formation from the central regions of active galactic nuclei, as well as from the relativistic jets emanating from them. Therefore, we have made several estimates concerning this class of objects. Below we will use the parameters of both the Event Horizon Telescope EHT~\cite{EHT} (signal accumulation time $t_{\rm obs} = 8$ hours, observation frequency $\nu = 345$ GHz) and the MOJAVE program~\cite{MOJAVEXVII} ($t_{\rm obs} = 30$ minutes, $\nu = 15.3$ GHz).

\begin{figure*}
\begin{minipage}{0.45\linewidth}
		\center{\includegraphics[width=1\linewidth]{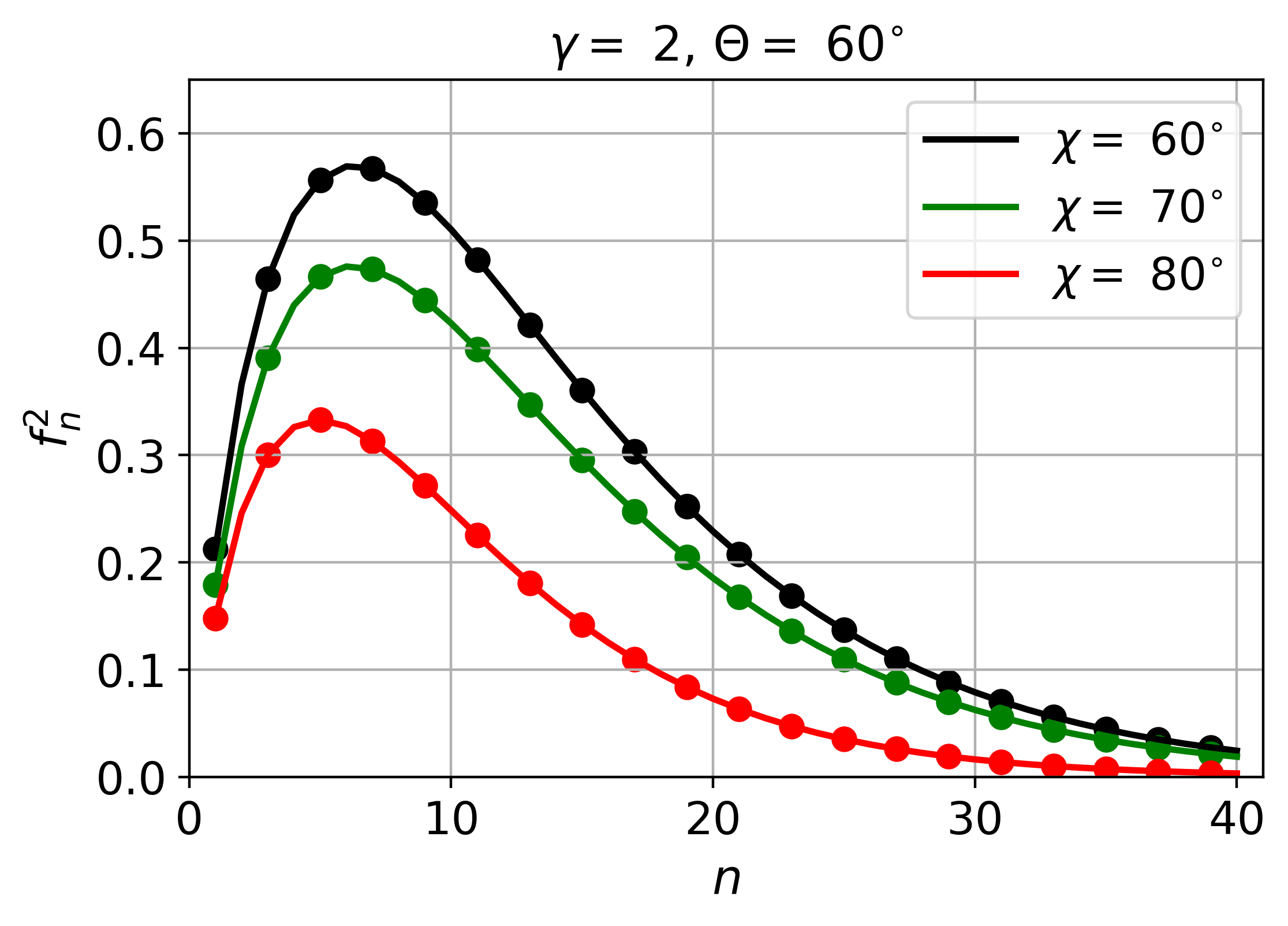} }
	\end{minipage}
	\hfill
	\begin{minipage}{0.45\linewidth}
		\center{\includegraphics[width=1\linewidth]{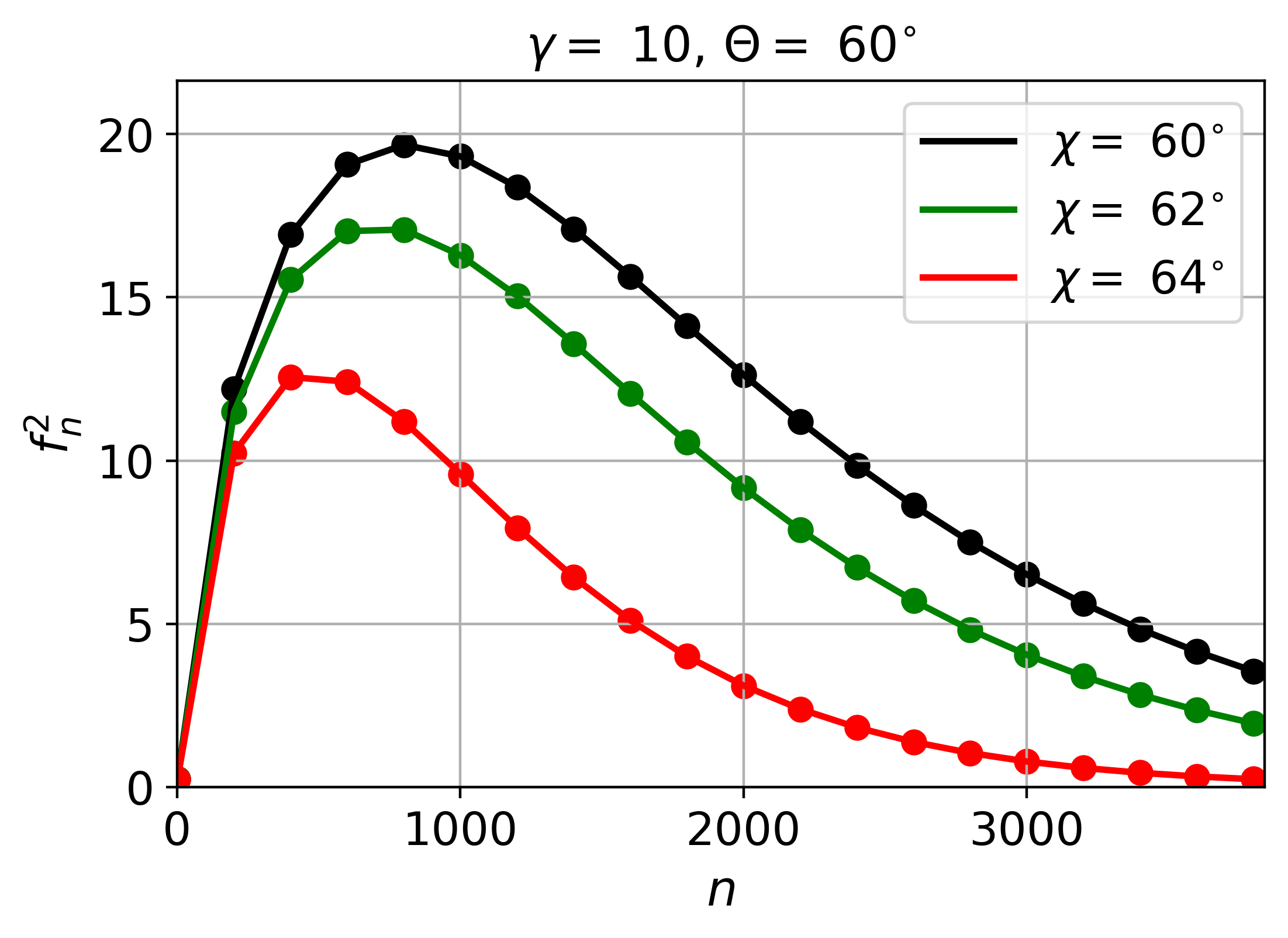}  }
	\end{minipage}
\caption{Comparison of the averaged synchrotron radiation spectra for various pitch angles $\chi$, 
obtained using the analytical expressions (\ref{eq:En})--(\ref{pn}) for the discrete 
spectrum (points) and the spectrum calculated by direct Fourier transform of the signal 
${\bf E}(r,t)$ (\ref{E}) (solid lines) for $\gamma = 2$ (left) and $\gamma = 10$ (right).
}
\label{fig2}
\end{figure*}

In what follows, the characteristic lengths $r$ will be normalized to the so-called light cylinder radius
\begin{equation}
R_{\rm L} = \frac{c}{\Omega},
\label{RL}
\end{equation}
where $\Omega$ is the angular velocity of the black hole or relativistic jet. According to numerous calculations (see, e.g.,~\cite{RL1}), the light cylinder should be approximately ten times the size of the black hole gravitational radius $r_{\rm g} = 2GM/c^2$. For M87, the corresponding value $\Omega \approx 10^{-6}$ s$^{-1}$ (which will be used below) was obtained directly from observations~\cite{Mertens} and gives the light cylinder radius of about $R_{\rm L} \approx 3 \cdot 10^{16}\; \text{cm} \approx 10 r_{\rm g}$. As an estimate of the magnetic field magnitude on the light cylinder $B_{\rm L}$, we have chosen $B_{\rm L} \sim 10^{2}$ G, which corresponds to various extrapolations of observational data.

The Lorentz factors of the emitting particles $\gamma$ must be chosen based on the condition of correspondence to the observation frequencies. In what follows, we will assume $\gamma \sim 10$--$10^{2}$ for EHT and $\gamma \sim 10^{2}$--$10^{5}$ for MOJAVE.

First of all, for the total emission time of a single particle $t_{\rm tot}$ we obtain: 
\begin{eqnarray}
t_{\rm tot}^{\rm EHT} & \approx & \frac{r}{\gamma c} \approx 300 
\left(\frac{r}{10^{14} \text{cm}}\right) \left(\frac{\gamma}{10}\right)^{-1} \text{s}, \\
t_{\rm tot}^{\rm MOJAVE} & \approx & \frac{r}{\gamma c} \approx 300 
\left(\frac{r}{10^{16} \text{cm}}\right) \left(\frac{\gamma}{10^3}\right)^{-1} \text{s}.
\label{ttot}
\end{eqnarray}

As we see, for observations near the black hole horizon, this time is certainly less than the observation time $t_{\rm obs}$. However, in the case of relativistic jets, the emission time can be either greater or less than $t_{\rm obs}$. Accordingly, the spectra of the received radiation should behave differently.

It is important to note that up to now we have discussed the synchrotron radiation from individual particles. At the same time, when considering relativistic jets, we are rather dealing with a quasi-stationary, continuous flow. In such a case, the observation time no longer determines the shape of the spectrum, and the corresponding parameter is actually the spatial resolution of the telescope. Indeed, in the stationary approximation, it is precisely the size $\Delta r$ of the region unresolved by the telescope that determines the segment of particle trajectories that mostly contributes to the radiation observed at a given point. Thus, the time $t_{\rm tot}$ must be compared not with $t_{\rm obs}$, but with the quantity $\Delta r / c$, for which  a simple estimate can be made. The angular resolution of modern VLBI networks is about one milliarcsecond. Taking the parameters for M87, we obtain a spatial resolution $\delta r \approx 10 - 100 \; r_{\rm g}$, which gives $\Delta r / c \approx 10^7 \: \text{s}$, which is many orders of magnitude greater than the characteristic times $t_{\rm tot}$. Since M87 is relatively close, this estimate is actually a lower bound and we find that for relativistic jets we always see radiation from the entire particle trajectory. We also note that we have deliberately set aside the question of the influence of a nontrivial pitch angle and energy distribution functions on the observed spectra. This question will be considered in Section~\ref{sec:ansamble}.

\begin{figure*}
\begin{minipage}{0.45\linewidth}
		\center{\includegraphics[width=1\linewidth]{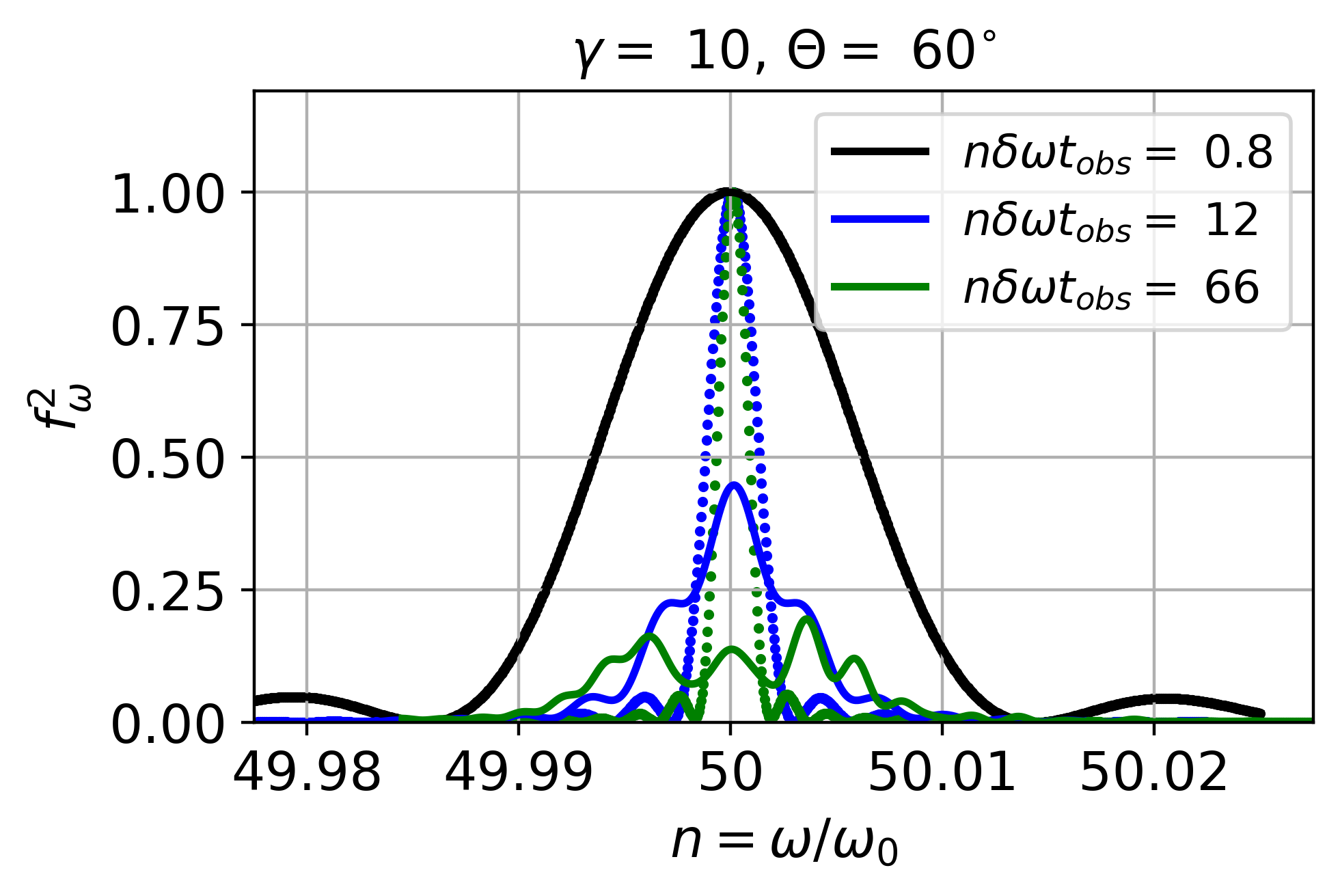} }
	\end{minipage}
	\hfill
	\begin{minipage}{0.45\linewidth}
		\center{\includegraphics[width=1\linewidth]{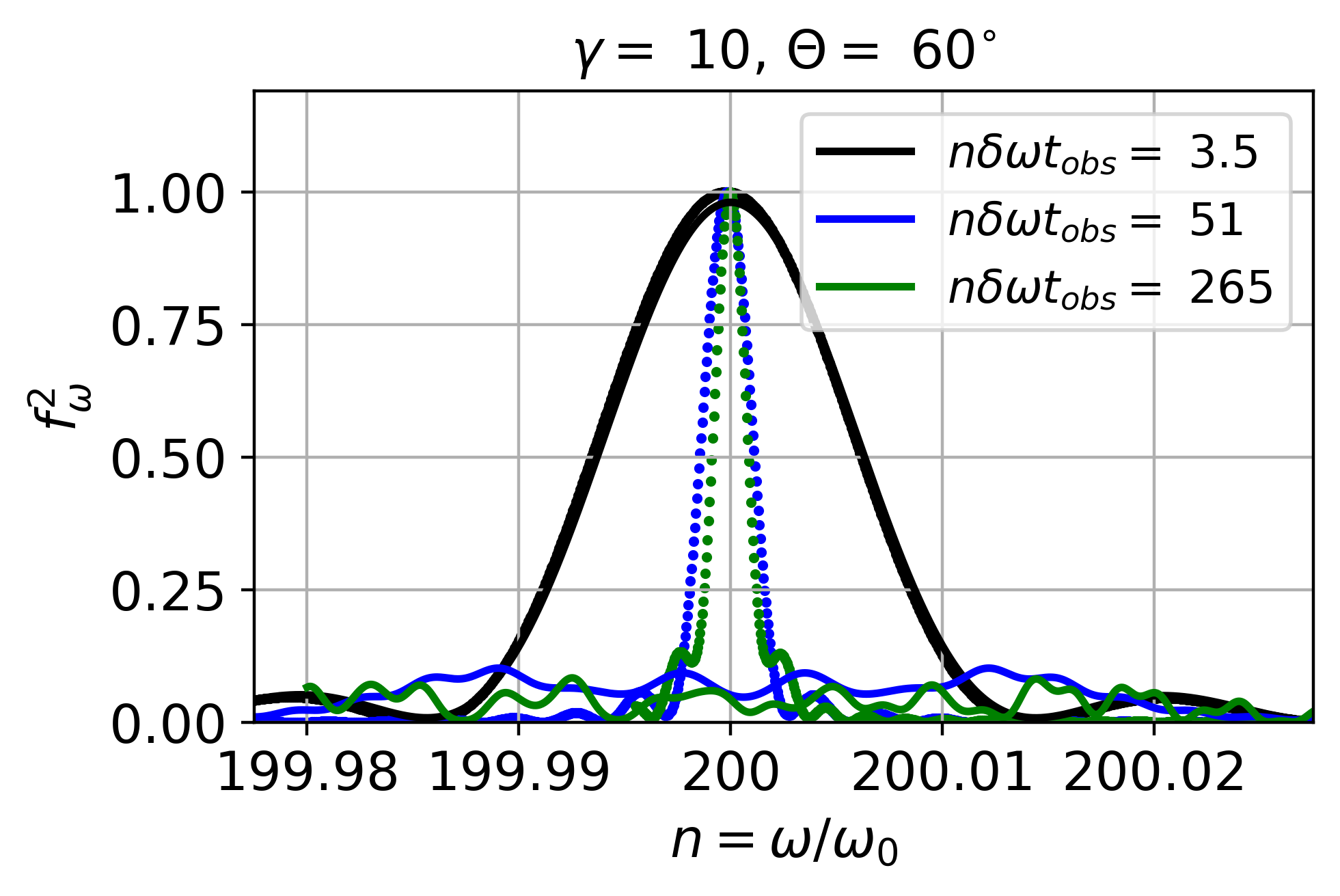}  }
	\end{minipage}
\caption{ Spectra in a constant (dashed) and non-uniform (solid lines) magnetic field  for the particle Lorentz-factor $\gamma = 10$ for various values of $\delta\omega \, t_{\rm obs}$ near the harmonics $n =50$ (left) and $n = 200$ (right). }
\label{fig3}
\end{figure*}

Let us now return to the problem of the radiation spectrum of a single particle, which is important from a theoretical perspective. Due to the "uncertainty relation" $\Delta \omega \Delta t \approx 1/2$, the finite observation time $\Delta t = t_{\rm obs}$, even for a uniform magnetic field, should lead to the transformation of the discrete spectrum with frequencies $n\omega_{0}$ ($\omega_{0} \approx \omega^{*}_{B}$) to the set of peaks with a characteristic relative frequency broadening 
\begin{equation}
\frac{\Delta \omega}{\omega_{0}} \approx \frac{1}{2\,t_{\rm obs}\,\omega_{B}}
\approx 10^{-9} \left(\frac{\gamma}{10^{2}}\right) 
\left(\frac{t_{\rm obs}}{1 \, \text{h}}\right)^{-1}
\left(\frac{B}{1 {\, \text{G}}}\right)^{-1}.
\label{Delta}
\end{equation}
As we see, in all the considered cases, such broadening remains sufficiently small, so there will be no peak overlap. At the same time, this example can be used to verify the adequacy of our approach, since in the study of the distortion of the synchrotron radiation spectrum, it will be sufficient for us to compare two continuous spectra rather than a continuous spectrum in a non-uniform magnetic field and a discrete spectrum in a uniform field.

Figure~\ref{fig2} shows the synchrotron radiation spectra for various pitch angles $\chi$ close to the observation angle $\Theta$ ($\delta \chi \sim 1/\gamma$). The points denote the result obtained with the analytical expressions (\ref{eq:En})--(\ref{pn}) for the discrete spectrum, while the solid lines represent the spectrum determined by direct Fourier transform of the signal ${\bf E}(r,t)$ (\ref{E}) and subsequent integration of the square of its amplitude over frequency $\omega$ near the corresponding harmonic $n\omega_{0}$. As we see, these spectra exactly match each other.

Another important circumstance is associated with the finite observation time $t_{\rm obs}$. If during this time the frequency change $\delta\omega$ is sufficiently small, so that $\delta\omega \, t_{\rm obs} \ll 1$, the radiation spectrum obtained within the assumption of a uniform magnetic field cannot be significantly distorted. The changes discussed below will occur only in the case of the opposite inequality $\delta\omega \, t_{\rm obs} > 1$.

Let us now estimate the magnitude of $\delta\omega \, t_{\rm obs}$. According to (\ref{omega}), we have $\delta\omega_{0} \approx \omega_{0} \, \delta r/r$, with $\delta r \approx c t_{\rm obs}$. As a result, we obtain
\begin{eqnarray}
\delta\omega \, t_{\rm obs} & \approx & \frac{\omega_{0} c}{r}t^{2}_{\rm obs}
\label{DD} \\
& \approx & 2 \times 10^{13}\left(\frac{r}{10^{16} \, \text{cm}}\right)^{-1} 
\left(\frac{\nu}{100 \, \text{GHz}}\right) 
\left(\frac{t_{\rm obs}}{1 {\, \text{h}}}\right)^{2}.
\nonumber
\end{eqnarray}
That is, even for $t_{\rm obs} \sim$ 1 s and $\nu =$ 1 GHz, this value remains much greater than unity up to distances $r > 10^{20}$ cm. Accordingly, at the harmonic number $n$ it will be even $n$ times larger. Thus, under real astrophysical conditions, one can expect a significant change in the synchrotron radiation spectrum.

Let us now see what actually happens to the synchrotron radiation spectrum as the value of $n \delta\omega t_{\rm obs}$ increases.
Figure~\ref{fig3} shows the spectra in uniform and inhomogeneous magnetic fields for various values of $n \delta\omega t_{\rm obs}$ at $\gamma = 10$ near the harmonics $n =50$ (left) and $n = 200$ (right). For each pair of values $n \delta\omega t_{\rm obs}$, they are normalized to the maximum value of the spectral density in a constant field. In all cases, the observation time $t_{\rm obs}$ was much less than the emission time $t_{\rm tot}$. Moreover, here and below, the frequency $\omega_{0}$ for the constant field was chosen as the arithmetic mean of the corresponding frequencies in the non-uniform field for the initial and final observation times.

As we can see, for $n \delta\omega t_{\rm obs} < 1$ the spectra in constant and non-uniform fields do indeed differ little from each other. However, as the observation time $t_{\rm obs}$ increases (and, consequently, $n \delta\omega t_{\rm obs} \propto t_{\rm obs}^2$), the widths of the peaks start to depend not on natural broadening, but on the real change in the fundamental frequency $\delta \omega \propto t_{\rm obs}$. Therefore, unlike in  the case of a constant field, where an increase in the observation time $t_{\rm obs}$ leads to a decrease in the spectral line widths $\Delta\omega$ (which do not depend on the harmonic number $n$), in a non-uniform field the widths of the peaks increase. In other words, these small-scale distortions are not related to the so-called spectral leakage effect, which inevitably arises in short-time Fourier transform. The broadening of spectral lines associated with this effect must repeat the spectrum of the chosen window function (in our case, rectangular), as seen in Fig.~\ref{fig3} for a constant field, and cannot depend on the harmonic number, unlike the distortions in the case of a non-uniform magnetic field.

However, as it turned out, such a strong difference between the spectra in inhomogeneous and uniform magnetic fields concerns only their small-scale structure. The integral energy flux in each harmonic (i.e., in the frequency interval \mbox{$\omega_{n} - 0.5 \, \omega_{0} < \omega < \omega_{n} + 0.5 \, \omega_{0}$)} under the condition $t_{\rm obs} \ll t_{\rm tot}$ in all cases coincides with the energy flux in the same harmonic for a constant magnetic field within 1\% accuracy. It should be noted that the constant and non-uniform magnetic fields were chosen such that the total radiation energy during the observation time $t_{\rm obs}$ was the same in both cases. In this case, from Plancherel's theorem \footnote{The Fourier transform $F$ is linear and bijective from the space $L_{2}\left(-\infty,+\infty\right)$ onto itself, and for any element $\phi \in L_2\left(-\infty,+\infty\right)$ the equality $||F\left[\phi\right]||_2 = ||\phi||_2$ holds.} \cite{Kudryavtsev}, it follows that the integrals of their Fourier transforms taken over the entire frequency range must also coincide, which allows a correct comparison of the intensities of the corresponding harmonics.

As the broadening of the $n$-th harmonic in a non-uniform field due to the broadening of the fundamental frequency $\omega_{0}$ by $\delta \omega_{0}$ should be $n$ times larger
\begin{equation}
\delta \omega_{n} \approx n \frac{\omega_{0} c}{r} \, t_{\rm obs},
\label{ndelta}
\end{equation}
the possibility to observe significant distortions in the averaged spectrum should be considered. Such broadening is clearly seen in comparison of the harmonics $n = 50$ and $n = 200$ in Fig.~\ref{fig3}. As a result, for $\gamma \gg 1$, when $n \sim \gamma^3$, the relative broadening should be of the order
\begin{equation}
\frac{\delta \omega_{n}}{\omega_{0}} \sim \frac{c t_{\rm obs}}{r}\,\gamma^{3} \sim 10 \, 
\left(\frac{\gamma}{10}\right)^{3} 
\left(\frac{r}{10^{16} \, \text{cm}}\right)^{-1} 
\left(\frac{t_{\rm obs}}{1 {\, \text{h}}}\right).
\label{ndel}
\end{equation}
Consequently, in the region of the synchrotron spectrum maximum, there may be a significant overlap of the peaks corresponding to different harmonic numbers $n$. This, in turn, could lead to a significant change in the averaged spectrum.

\section{Distortion of the averaged spectrum}
\noindent

In Section 3, in order to compute the synchrotron radiation spectrum, we directly determined the time dependence of the electric field \(\mathbf{E}(t)\) at the observation point and performed the corresponding Fourier transform. Although this approach is the most accurate and is applicable to any configuration of external magnetic fields, it suffers from a significant problem: in order to obtain the averaged spectrum one must first calculate it with high temporal resolution, which makes the method extremely computationally intensive. Its application is particularly difficult in the most interesting ultrarelativistic case, since the required resolution in both time and frequency domains grows as the cube of the gamma factor ($\Delta t/t_{\rm obs} \sim 1/(\omega_{\rm B}\gamma^{3}t_{\rm obs}) \sim \gamma^{-3}; \Delta \omega/\omega_{\rm max} \sim \omega_{0}/\omega_{\rm max} \sim \gamma^{-3}$). Hence, one should develop an approximate semi-analytical method to find the averaged spectrum.

To do this, consider the Fourier spectrum of the vector potential in the case of a varying magnetic field. Being no longer discrete, it will be determined by the following expression:
\begin{equation}
\mathbf{A}{\omega} = \frac{e^{i k R_0}}{c R_0} \int{t_{\rm{obs}}} e \mathbf{v}(t) e^{i[\omega t - \mathbf{k} \mathbf{r}(t)]} dt,
\label{eq:A_expr}
\end{equation}
where $R_0$ is the distance from the radiation source to the observer, and the integration is performed over the entire observation time. It turns out that in the weakly inhomogeneous magnetic field approximation \mbox{$(c/\omega_B)\cdot(\nabla\omega_B/\omega_B)\ll1$,} the evaluation of the integral \eqref{eq:A_expr} can be significantly simplified by reducing the problem to the well-known case of a constant magnetic field.
 
Indeed, in this approximation the particle's motion can be represented as a set of turns, on the scale of which the variation of the magnetic field can be neglected. The integration over the entire observation time can then be replaced by a sum of integrals over "quasi-periods", within each of which the magnetic field may be considered constant. Thus, the expression for the resulting electric field reduces to a sum of terms of the form \eqref{eq:En} with the formal replacement of the harmonic number \(n\) by the ratio \(\omega / \omega_0^k\) and the addition of a phase factor 
$\exp [i \omega(t^k - r^k\cos{\Theta}/c)$]:
\begin{eqnarray}
\mathbf{E}_{\omega}(\Theta, r)   =  \sum_{k = k_{\min}}^{k_{\max}} \exp\left[i \omega\left(t^k - \frac{r^k\cos{\Theta}}{c}\right)\right] \mathbf{E}^k_{\omega}, \\
\mathbf{E}^k_{\omega}  = \frac{2e}{cR_0}\,\frac{\omega}{\omega^k_0}\frac{\omega^{*k}_{B}\beta\sin\chi^k}{(1 - \beta\cos\chi^k\cos\Theta)^2} 
\nonumber \\
\times \left[\mathbf{l}_{1}J'_{\omega/\omega_0^k}(z_{\omega}^k) - i\mathbf{l}_{2}\frac{\cos\Theta - \beta\cos\chi^k}{\beta\sin\chi^k\sin\Theta}J_{\omega / \omega_0^k}(z^k_{\omega})\right],
\label{eq:E_omega}
\end{eqnarray}
where
\begin{equation}
    z_{\omega}^k = \frac{\omega}{\omega_0^k} \frac{\beta \sin{\chi^k}\sin{\Theta}}{1 - \beta \cos{\chi^k}\cos{\Theta}},
\end{equation}
and the index \(k\) numerates the turns into which the particle's motion is divided. The specific geometry of the magnetic field will, in turn, determine the dependence of all spatially-dependent quantities on the index \(k\), as well as the relation between the index \(k\) and time.

The radiation intensity in turn can be calculated as follows:
\begin{eqnarray}
&&    {\rm d}I   =  \frac{c}{2 \pi} \left|\mathbf{E_\omega} \right|^2 R_0^2 {\rm d}\Omega
    \\
&&      =  \frac{cR_0^2}{2\pi}\left|\sum_{k = k_{min}}^{k_{max}} \exp\left[i \omega \left(t^k - \frac{r^k \cos{\Theta}}{c}\right)\right] \mathbf{E}^k_\omega \right|^2 d\Omega.
    \nonumber 
\end{eqnarray}
The expression for high frequencies \(\omega \gg \omega_0\) can be significantly simplified by averaging over a small frequency interval \(\omega_0 \ll \Delta\omega \ll \omega\):
\begin{align}
& \left\langle \frac{{\rm d}I}{{\rm d}\Omega} \right\rangle_{\Delta \omega} 
= \frac{cR_0^2}{2\pi}\sum_{n,m = k_{\min}}^{k_{\max}}
\left\langle \exp\left[i \omega \left(t^n - \frac{r^n \cos{\Theta}}{c}\right)\right] \right.
\nonumber \\
& \left. \times \exp\left[-i \omega \left(t^m - \frac{r^m \cos{\Theta}}{c}\right)\right]\right\rangle_{\Delta \omega} \mathbf{E}^n_\omega \mathbf{E}^m_\omega.
\end{align}
Taking into account that \((t_n - t_m) \geq 1/\omega_0\) (and similarly \(r^m/c - r^n/c\)) for \(n\neq m\), we find that for sufficiently large 
$\Delta \omega$, $\left\langle e^{i \omega (t^n - r^n \cos{\Theta}/c)} e^{-i \omega (t^m - r^m \cos{\Theta}/c)}\right\rangle_{\Delta \omega} \approx \delta_{n, m}$. Then, for the averaged angular intensity distribution we obtain:
\begin{equation}
    \left\langle \frac{{\rm d}I}{{\rm d}\Omega} \right\rangle_{\Delta \omega} = \frac{cR_0^2}{2\pi}\sum_{k = k_{\min}}^{k_{\max}} \left|E_\omega^k\right|^2, 
    \label{eq:dI/dOmega}
\end{equation}
where \(E_\omega^k\) is defined by the expression \eqref{eq:E_omega}.

\begin{figure*}
\centering
\includegraphics[scale=0.43]{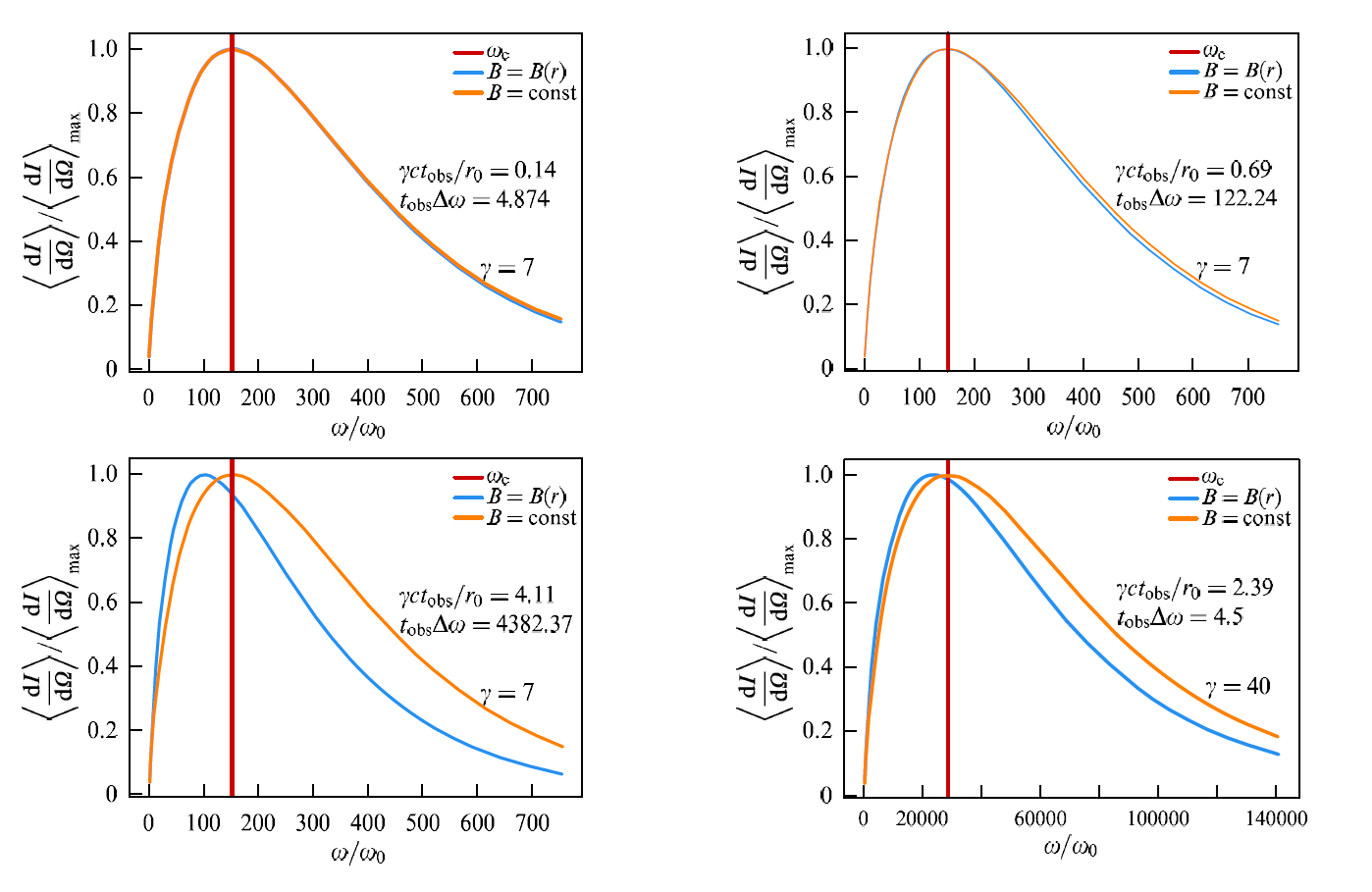}
\caption{Comparison of the averaged synchrotron spectrum for the cases of monopole and uniform magnetic fields.}
\label{fig:monopole_spectra}
\end{figure*}

Note that the two terms in expression \eqref{eq:E_omega} in fact determine the intensities of the two linear polarization components. Expressions \eqref{eq:E_omega} and \eqref{eq:dI/dOmega} generalize relations \eqref{eq:En} and \eqref{pn} to the case of a weakly inhomogeneous magnetic field.

Now it is possible to obtain the criterion for the change in the averaged intensity spectrum. Indeed, as can be seen from \eqref{eq:E_omega} and \eqref{eq:dI/dOmega}, the spectrum in a non-uniform field is composed of expressions similar to those for a constant magnetic field, but with parameters varying within certain ranges. These expressions mostly depend on the particle's pitch angle \(\chi_k\); a variation of this angle by an amount on the order of \(1/\gamma\) leads to a significant change in the arguments of the corresponding Bessel functions.

Thus, we conclude that for sufficiently large gamma factors, in the first approximation one can neglect the variations of the magnetic field amplitude along the trajectory and consider only the associated change in the particle's pitch angle. Therefore, a significant change in the averaged synchrotron spectrum can be expected when the condition \(\gamma  \delta \chi \gtrsim 1\) is satisfied, where \(\delta \chi\) is the change in the particle's pitch angle over the observation time \(t_{\mathrm{obs}}\)  (see Fig. \ref{fig:monopole_spectra}). This means that for observation times much shorter than the time it takes the observer to leave the radiation beam, the distortions of the averaged synchrotron spectrum can be neglected.
 
\section{Synchrotron radiation of the particle ensemble}
\label{sec:ansamble}
\noindent

In real astrophysical problems involving relativistic jets, one always deals with an ensemble of radiating particles possessing certain energy and (initial) pitch angle distribution functions \cite{Baldwin1977, Comisso&Sironi2019, Comisso&Jiang2023}. At the same time, as is well known, the result depends significantly on whether we consider a steady particle flow or a radiating plasma cloud moving toward the observer~\cite{Sazonov, Ginzburg}. Here we consider only the first case, although the other problem statement is also relevant in the case of jets from active galactic nuclei.

If the particles radiate incoherently, the resulting intensity spectrum can be computed as a sum of the radiation spectra of individual particles, taking into account their distribution function:
\begin{equation}
        \left\langle \frac{dI}{d\Omega} \right\rangle_{\Delta \omega}^{\mathrm{tot}} = \int \int d\chi_0 d \epsilon f(\chi_0, \epsilon) \left\langle \frac{dI}{d\Omega} \right\rangle_{\Delta \omega} (\chi_0, \epsilon).
    \label{eq:dI/dOmega_total}
\end{equation}
To simplify the analysis, we first of all consider a monoenergetic particle distribution
\begin{equation}
f(\chi_0, \epsilon) = f_{\chi}(\chi_0) \delta(\epsilon - \epsilon_0).
\end{equation}

Here \(\epsilon_0\) is the particle energy (which is constant in the approximation of small radiation losses), and \(\chi_0\) is its initial pitch angle. Of course, in nature there always exists some non-trivial particle energy distribution. However, as will be shown in this chapter, this approximation proves sufficient for studying the influence of magnetic field inhomogeneity on the averaged synchrotron spectrum. The key observation is that, despite the freedom in choosing the distribution function \(f_\chi (\chi_0)\), it is precisely the gamma factor of the particles, corresponding to the energy \(\epsilon_0\), that determines the shape of the final spectrum. Indeed, in the ultrarelativistic regime \(\gamma \gg 1\), the largest contribution to the resulting spectral density of radiation comes from the particles whose pitch angles differ from the observation angle \(\Theta\) by no more than \(\sim 1/\gamma\). Thus, the integral over the pitch angle in expression \eqref{eq:dI/dOmega_total} accumulates only in a small neighborhood of the angle \(\Theta\), which allows one to approximate any sufficiently smooth distribution \(f_\chi(\chi_0)\) as uniform:
\begin{eqnarray}
     \left\langle \frac{dI}{d\Omega} \right\rangle_{\Delta \omega}^{\mathrm{tot}} = \int_0^{\pi/2} d\chi_0 f(\chi_0) \left\langle \frac{dI}{d\Omega} \right\rangle_{\Delta \omega} (\chi_0) 
     \nonumber  \\
     \approx  \frac{2}{\pi} \int_{0}^{\pi/2} d\chi_0 \left\langle \frac{dI}{d\Omega} \right\rangle_{\Delta \omega} (\chi_0).
     \label{eq:average}
\end{eqnarray}
Now using expression \eqref{eq:dI/dOmega} for \(\left\langle {\rm d}I{{\rm d}\Omega} \right\rangle_{\Delta \omega}\) and the classical ultrarelativistic approximation \(\chi_k \approx \Theta\), we obtain:
\begin{eqnarray}
&&   \left\langle \frac{dI}{d\Omega} \right\rangle_{\Delta \omega}^{\mathrm{tot}} = 
   \frac{cR_0^2}{2\pi}\sum_{k = k_{\min}}^{k_{\max}} \int_0^{\pi/2} d\chi_0 \left|E_\omega^k\right|^2 (\omega^k_0, \omega^{*k}_B, \chi^k) 
   \nonumber \\
 &&  \approx \frac{cR_0^2}{2\pi}\sum_{k = k_{\min}}^{k_{\max}} \int_0^{\pi/2} d\chi_k \frac{d\chi_0}{d\chi_k}\left|E_\omega^k\right|^2 (\omega_0, \omega^{*}_B, \chi^k) 
   \nonumber \\
 &&  \approx \frac{cR_0^2}{2\pi}\sum_{k = k_{\min}}^{k_{\max}} \int_0^{\pi/2} d\chi_k\left|E_\omega^k\right|^2 (\omega_0, \omega^{*}_B, \chi^k) 
   \nonumber \\
&&   \propto \int_0^{\pi/2} d\chi_k\left|E_\omega^k\right|^2 (\omega_0, \omega^{*}_B, \chi^k) \propto {\cal J}(\omega),
\end{eqnarray}
where \({\cal J}(\omega)\) is defined by the expression \eqref{eq:integrated_spectrum}.
 
\begin{figure*}[h]
\begin{minipage}{0.45\linewidth}
\center{\includegraphics[width=1\linewidth]{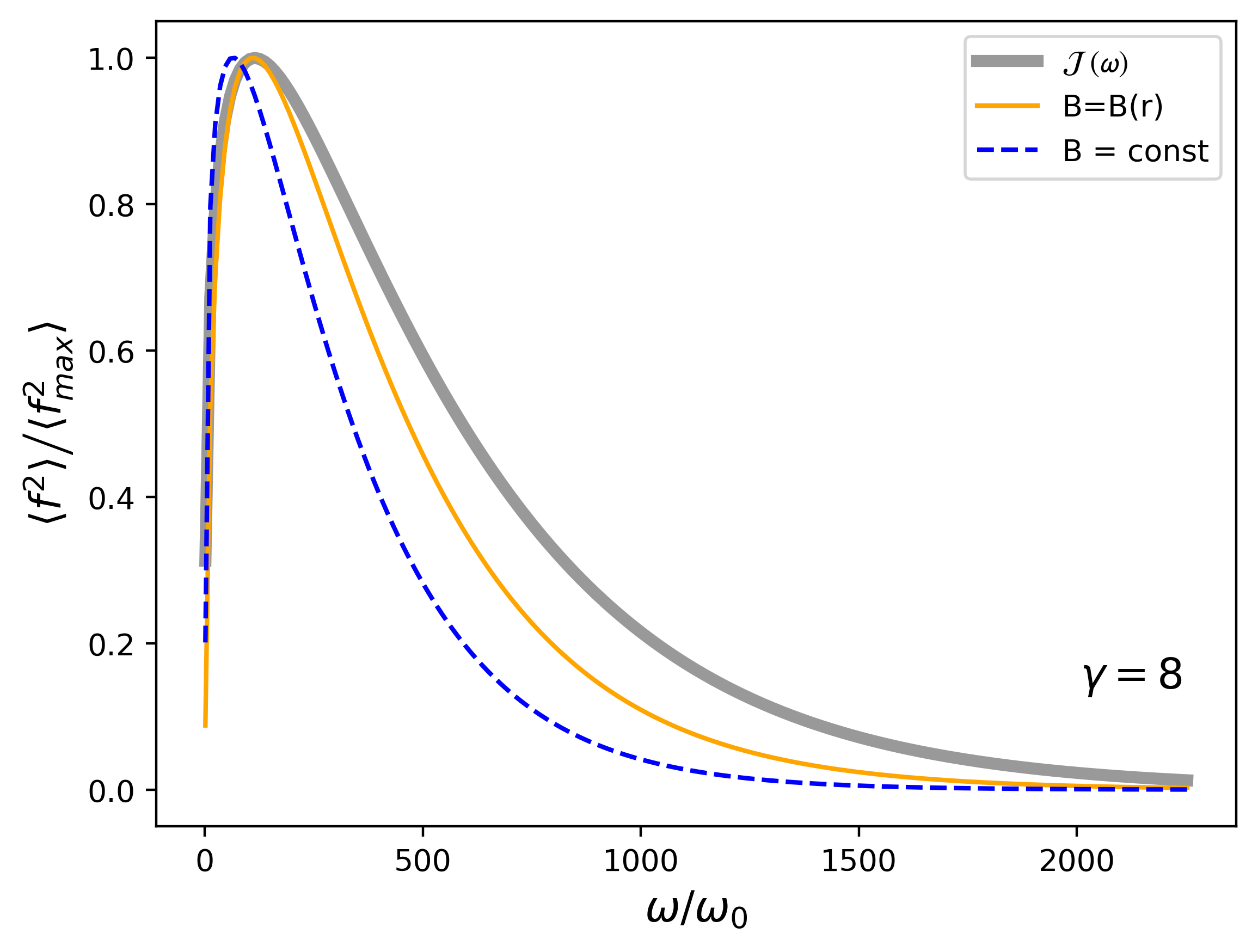} }
\end{minipage}
\hfill
\begin{minipage}{0.45\linewidth}
\center{\includegraphics[width=1\linewidth]{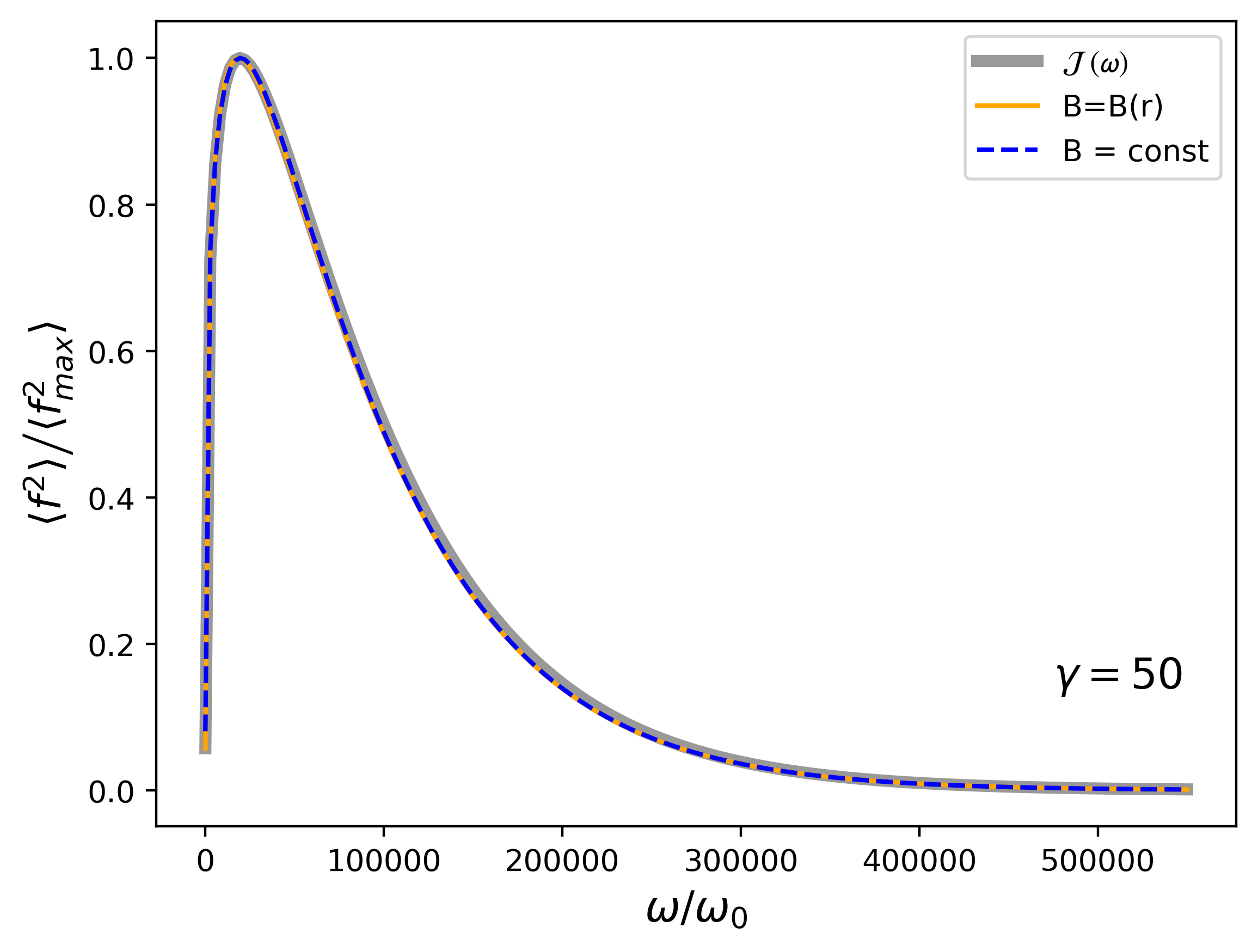}  }
\end{minipage}
\caption{Comparison of averaged synchrotron spectra of an ensemble of particles with a non-degenerate angular distribution function for two different \(\gamma\)-factors.}
\label{fig:many_particles}
\end{figure*}

Thus, when the particle gamma factors are much greater than unity, the presence of a non-degenerate particle distribution in pitch angle (which is always present in real astrophysical problems), regardless of the geometry of the large-scale magnetic field, leads to the fact that the spectral shape coincides with the classical expression for the total spectral power loss \eqref{eq:dI/dOmega}. Here it is worth emphasizing once again that we are still dealing with the radiation intensity received by a distant observer, but the averaging over the pitch angle becomes equivalent to the integration over the observation angle. However, in the subrelativistic regime \(\gamma \sim (5 -10)\) the specific form of the angular distribution function of particles begins to significantly influence the value of the integral in expression \eqref{eq:average}. Therefore, the corresponding distortions, related to the inhomogeneity of the magnetic field, lead to distortions in the radiation spectrum (Fig. \ref{fig:monopole_spectra}, \ref{fig:many_particles}).

\section*{Conclusions}
\noindent

Therefore, a detailed analysis of synchrotron radiation from \textit{individual} particles has demonstrated that inhomogeneities in the magnetic field may result in changes in both small- and large-scale features of the synchrotron spectrum. Small-scale distortions, or broadening of spectral lines, are associated not with the finite duration of the observation, but rather with changes in the cyclotron frequency along the particle trajectory, and occur under the condition $n \delta\,\omega\,t_{\rm obs} \gg 1$, which is satisfied for most astrophysical sources. These distortions, however, are of purely theoretical interest, as they disappear completely upon averaging the spectrum over small scales due to the finite spectral resolution of radio telescopes. A significant change in the averaged spectrum is possible, although it occurs only for observation times that are longer than the particle emission time in the direction of the observer. This change is primarily due to the variation in the pitch angle of the particle as it travels through an inhomogeneous magnetic field.

With regard to real astrophysical sources, where the radiating particles have non-degenerate energy and pitch angle distribution functions, the inhomogeneity of the large-scale magnetic field plays an even less significant role in the formation of the averaged spectrum. Indeed, in the ultrarelativistic regime $\gamma \gg 1$ that is most relevant for astrophysical applications, averaging over the angular distribution function is equivalent to integrating over the observation angle, leading to the classical expression for the total spectral losses. Conversely, in the subrelativistic regime at $\gamma \sim (5-10)$, some deviations from the averaged spectrum may be expected. Nevertheless, a detailed investigation of this phenomenon is beyond the scope of this research.

The authors thank E.E.~Nokhrina and A.P.~Lobanov for valuable discussions. This study was supported by the Russian Science Foundation, grant no. 24-22-00120.

\section{Appendix. The trajectory of the individual particle in the monopole magnetic field}

For the monopole magnetic field, an exact expression is known for the trajectory of the charged particle in the spherical coordinate system
\begin{align}
& \theta  =  \arctan \left( \frac{r_0 v\sin{\chi_0}}{\omega_{B} R_{\rm L}^2} \right), 
\label{eqofmot_theta} \\
& \varphi =  \varphi_0 + \sqrt{1 + \left(\frac{\omega_{B}R_{\rm L}^2}{r_0 v\sin{\chi_0}}\right)^2} \arctan \left(\frac{tv^2 + r_0 v\cos{\chi_0}}{r_0 v\sin{\chi_0}}\right), 
\label{eqofmot_phi} \\
& r =  \sqrt{r_0^2 + 2r_0 t v\cos{\chi_0} + t^2 v^2}.
\label{eqofmot_r}
\end{align}
In this case, the reference frame is selected such that the effective angular momentum
\begin{equation}
{\bf J} = m_e\left[{\bf r} \times {\bf v}\right] +  \frac{eB_{\rm L}R_{\rm L}^2}{c}\frac{\bf r}{r},
\end{equation}
which serves as an integral of motion, is aligned with the $z$-axis. As a result, the charged particle follows a conical spiral trajectory
\begin{equation}
r \left(\varphi \right) = \frac{r_0 v\sin{\chi_0}}{v \Bigg|\cos{\left(\left[1 + \left(\frac{\omega_{B}R_{\rm L}^2}{r_0 v\sin{\chi_0}}\right)^2\right]^{-1/2}\varphi\right)}\Bigg|}.
\end{equation}
From the pitch of this spiral, it is possible to derive an exact expression for the angular velocity of the relativistic particle in the circular orbit in the $xy$-plane of the monopole field.
\begin{eqnarray}
&& \omega_{B}^{*} = \\
&& = \omega_B \frac{R_{\rm L}^2}{r^2} \left[ 1 - \frac{2\pi r_0 v\sin{\chi_0}}{\omega_B R_{\rm L}^2} \sqrt{\frac{r^2v^2}{r_0^2\left(v\sin{\chi_0}\right)^2} - 1}\right].
\nonumber
\end{eqnarray}
If the magnetic field is large-scale (i.e., $R_{\rm L} \gg r_{B}$, where $r_{B}$ is the gyroradius, which also varies with distance $r$ in this case), the frequency can be expressed simply
\begin{equation}
\omega_B^{*} \approx \omega_B \frac{R_{\rm L}^2}{r^2}.
\end{equation}

However, due to the Doppler shift, the frequency observed by an external observer, $\omega_0^{*}$, is of particular interest. Its approximate value can be derived directly from the retarded time $t^{\prime} = t + {\bf r}\left(t^{\prime}\right) {\bf n}/c - R_0/c$, where ${\bf n}=\left(0,\sin \Theta, \cos \Theta\right)$ is a unit vector along the line of sight, $R_0$ is the distance from the source to the observer.

Indeed, when considering the quasi-period of pulses received by the observer $T = t\left(\varphi + 2\pi\right) - t\left(\varphi\right)$ it is possible to obtain
\begin{align}
& T = T^{\prime} \\
&  + \frac{\left[r\left(\varphi\right) - r\left(\varphi + 2\pi\right)\right]}{c}\left(\sin \Theta \sin \theta \sin \varphi + \cos \Theta \cos \theta\right),
\nonumber
\end{align}
where $T^{\prime} = 2\pi/\omega_B^{*}$. Then, by substituting the equations of the trajectory \eqref{eqofmot_theta}--\eqref{eqofmot_r}, the expression for $\omega_0^{*}$ takes the following form:
\begin{align}
& \omega_0^{*} = 
\label{omega_0_expression} \\
&\frac{\omega_B^{*}}{1 - \beta \cos \chi \left(\cos \Theta + a \sin \Theta \sin \frac{\pi/2 - \chi}{a} - 2\pi a \cot \chi \cos \Theta\right)},
\nonumber
\end{align}
where the pitch angle $\chi$, as determined by conservation of the first adiabatic invariant \eqref{rsinchi}, depends on distance. It is worth noting that the expression for the frequency (\ref{omega_0_expression}) is obtained in the first approximation using a small value
\begin{equation}
a=\frac{r_0 v \sin \chi_0}{\omega_B R_{\rm L}^2},
\end{equation}
and for the selected parameters $B_{\rm L}=10^2~\text{G}$, and $R_{\rm L}=10^{16}~\text{cm}$, we have an estimate of $a \sim 10^{-12}$, which allows us to ignore the first-order term for the considered distance and represent the observed frequency as
\begin{equation}
\omega_0^{*} \approx \frac{\omega_B^{*}}{1 - \beta \cos \chi \cos \Theta},
\end{equation}
which is used in this study \eqref{omegar}. 

Furthermore, using the equations for the trajectory (\ref{eqofmot_theta})--(\ref{eqofmot_r}), we can obtain the dependence of the parameters $\omega_0^{*k}$, $\omega_B^{*k}$, and $\chi^k$ on the number of full turns of the spiral trajectory of particle $k$. These parameters are necessary to calculate the averaged spectrum using formulas (\ref{eq:E_omega}) and (\ref{eq:dI/dOmega}). The condition for a significant change in the spectrum, $\gamma\cdot \delta\chi \gtrsim 1$, is then expressed as $\gamma t_\mathrm{obs} / r_0 \gtrsim 1$. As shown in Fig.~\ref{fig:monopole_spectra}, the parameter $\gamma \delta \chi = \gamma t_\mathrm{obs}/r_0$ determines the degree of change in the averaged spectrum. In this case, the condition $t_\mathrm{obs}\Delta\omega \gg 1$ does not play a decisive role.

{}

\end{document}